\documentclass[a4paper,floatfix,amsmath,amssymb,prb,twocolumn]{revtex4-1}
\usepackage{amsmath}
\usepackage{booktabs}
\usepackage{graphicx}
\usepackage{epsf}
\usepackage{dcolumn}
\usepackage{bm}
\usepackage{color}
\usepackage{pstricks}
\usepackage{subfigure}
\usepackage{moreverb}
\usepackage{placeins}
\begin{document}

\title{Magneto-resistive property study of direct and indirect band gap thermoelectric Bi-Sb alloys}

\author{Diptasikha Das$^{1}$, K. Malik$^{1}$, S. Bandyopadhyay$^{1,2}$, D. Das$^3$, S. Chatterjee$^3$ and Aritra Banerjee$^{1,2,*}$\\ 
   $^1$Department of Physics, University of Calcutta, Kolkata 700 009, India\\ $^2$Center for Research in Nanoscience and Nanotechnology, University of Calcutta, JD-2, Sector-III, Saltlake, Kolkata 700 098, India\\
   $^3$UGC-DAE Consortium for Scientific Research, Kolkata Centre, Sector III, LB-8, Salt Lake, Kolkata 700 098, India\\
   $^*$Author to whom correspondence should be addressed. \textit{Electronic mail}: arbphy@caluniv.ac.in}

\begin{abstract}
We report magneto-resistive properties of direct and indirect band gap Bismuth-Antimony (Bi-Sb) alloys. Band gap increases with magnetic field. Large positive magnetoresistance (MR) approaching to $400\%$ is observed. Low field MR experiences quadratic growth and at high field it follows a nearly linear behavior without sign of saturation. Carrier mobility extracted from low field MR data, depicts remarkable high value of around $5$ $m^2V^{-1}s^{-1}$. Correlation between MR and mobility is revealed. We demonstrate that the strong nearly linear 
MR at high field can be well understood by classical method, co-build by Parish and Littlewood. 
\end{abstract}

\maketitle

Among Thermoelectric (TE) materials, Bi-Sb alloys have received special attention due to their interesting physical properties.\cite{4} Band structures of Bi and Bi-Sb alloy are 
peculiar in nature and depend drastically on different physical parameters including applied magnetic field.\cite{2,3} From materials
perspective, efficiency of a TE material is determined by its dimensionless figure of merit $ZT$ = $\frac{S^2}{\rho\kappa}$, where S is
the Seebeck coefficient, ${\rho}$ the electrical resistivity and $\kappa$ the total thermal conductivity.\cite{4} Mobility of carrier and different scattering mechanisms 
control the transport properties.\cite{5} The details of the Fermi surface geometry, character of electron-electron (e-e) interactions
and carrier mobility are strongly influenced in presence of magnetic field, which in turn affects the transport properties
including resistivity.\cite{3,6} MR of a TE material therefore gives valuable insight into the mechanism affecting conductivity and hence is intimately related to the underlying physics towards obtaining materials with high
ZT values. Further, materials with large and non trivial MR provide great opportunities in magnetic sensor and memory applications.\cite{7,8}  Bi and Bi-rich alloys possess very large MR and are potential candidates for quantum transport and finite-size effects.\cite{9} Yang et al demonstrated that, the advantage of such systems over other magnetic field sensors is the ability to measure fields in different orientations relative to the device without the loss of sensitivity.\cite{9} In addition, Bi-Sb alloy system also exhibit the fascinating properties of Topological Insulator, which apart from possessing interesting electronic properties, have potential applications in quantum information processing, magneto-electronic devices etc. Bi is rather ``an old material'', but owing to its unique magneto-transport properties, it shows unrelenting interest among researchers.\cite{9,10,11,12,13} Among other important Bi based TE materials, magneto-resistive properties of $Bi_2Te_3$ and $Bi_2Se_3$ have been 
systematically studied.\cite{6,7,8,14,15,16,17,18,19,20} On the other hand, high field MR and quantum oscillations in topological insulator Bi-Sb alloys have also been extensively studied.\cite{3,21,22,23} But there is a lack of systematic investigation on magneto-transport properties of semiconducting Bi-Sb alloy, which is a remarkable n-type TE material.\cite{24,25,26,27} However, the general consensus regarding origin of giant and Linear MR (LMR) observed in these TE alloys is yet to be obtained. Also, the relation between the MR and carrier mobility has still not been revealed for Bi-Sb alloys.

  The band diagram of Bi-Sb alloy depicts that, Sb concentration causes transition of the material from direct band gap semiconductor to indirect band gap one and power factor (PF$=\frac{S^2}{\rho}$) of the former is higher.\cite{28,29} In the present work we report temperature and magnetic field dependent 
resistivity of a direct band gap $(Bi_{0.90}Sb_{0.10})$ and an indirect band gap $(Bi_{0.86}Sb_{0.14}­)$ Bi-Sb alloy. Giant non-saturating MR at low temperature has been reported in polycrystalline $Bi_{1-x}Sb_{x} (x=0.10, 0.14)$ samples. Direct band gap $Bi_{0.90}Sb_{0.10}$ alloy possesses higher MR. Very large MR of about $400\%$ reported here in polycrystalline Bi-Sb system demonstrates the high quality of samples and indicates it's potential application as magnetic memory device. Further, a correlation between mobility and MR in Bi-Sb system has also been pointed out.

  Polycrystalline $Bi_{0.90}Sb_{0.10}$ and $Bi_{0.86}Sb_{0.14}$ samples are grown from stoichiometric mixture of $99.999\%$ purity Bi
and Sb elements (Alfa Aesar, UK) by solid state reaction method, details of which can be obtained elsewhere.\cite{28} ${MR}$ is generally defined as: ${MR}$ = $\frac{\Delta\rho}{\rho}$=$\frac{\rho(H)-\rho(0)}{\rho(0)}$ where, $\rho(H)$ and
$\rho(0)$ are resistivities in presence and absence of magnetic field respectively. Resistivity and MR measurement are performed on rectangular samples in a Closed Cycle Refrigerator (CCR) based 15 T magnet supplied by Cryogenic 
Ltd., UK in the range 5 to 300 K, where the standard four-probe technique with silver paste cured at room temperature is used 
for electrical contacts.

  Temperature dependence of resistivity, both in absence and presence of magnetic field (B=8, 15 T) for $Bi_{1-x}­Sb_{x}$ (x${=}$0.10 and 0.14) 
samples are shown in Fig. 1. The $\rho$-T curve shows non-monotonic temperature dependence with maximum value of $\rho$ at the temperature T$_p$.
\FloatBarrier
\begin{figure}[h]
\begin{center}
\includegraphics[scale=0.4]{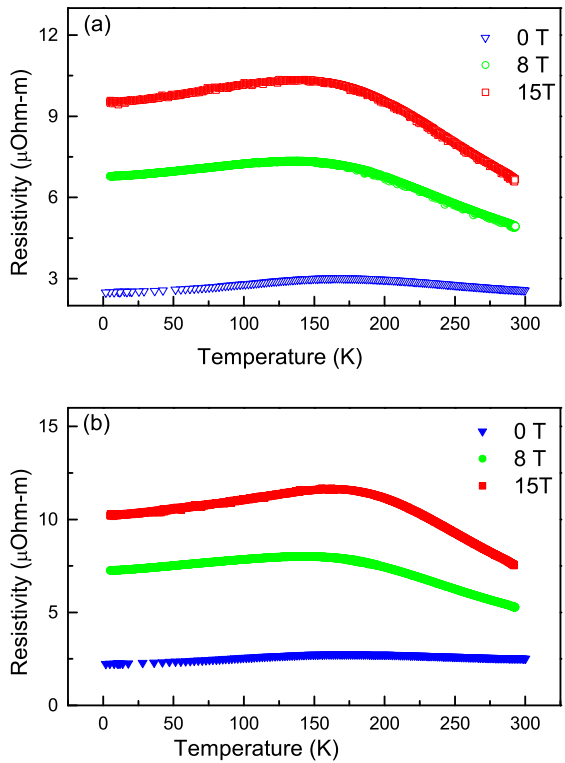}
\caption{(Color online) Temperature dependent electrical resistivity of (a) $Bi_{0.86}Sb_{0.14}$ and (b) $Bi_{0.90}Sb_{0.10}$ alloys at three different
magnetic fields (0T, 8T, 15T).}
\end{center}
\end{figure}
$\rho$ increases in presence of magnetic field for both the samples. In conventional metals Lorentz force induced by a magnetic
field bends the trajectory of an electron, which in turn gives rise to an increase in electrical resistance.\cite{30}
Band gap ($E_g$) of the samples have been estimated from the $\rho$(T) data both in absence and presence of
magnetic field. $E_g$ is measured from high temperature semiconducting region $(T > T_p)$ of $\rho$(T) data using the equation:
\begin{equation}
  \rho=\rho_{0} \exp(\frac{E_g}{2K_BT}),                
\end{equation}
where, $\rho_{0}$ is a constant. $E_g$ increases with magnetic field (Table-1) for both of the studied Bi-Sb samples. 
\begin{table}
\caption{Evolution of $E_g$ in presence of magnetic field as obtained for polycrystalline $Bi_{0.86}Sb_{0.14}$ and $Bi_{0.90}Sb_{0.10}$ alloys.}
\setlength{\tabcolsep}{12pt}
\centering
\begin{tabular}{ccc}

\hline
\hline
     \multicolumn{1}{c}{Magnetic Field (T)} & \multicolumn{2}{c}{\(E_{g}\) (meV)}\\
     
         & Bi$_{0.86}$Sb$_{0.14}$    &    Bi$_{0.90}$Sb$_{0.10}$   \\
    \hline
    0   & 9.66 & 7.77  \\
    
    8   & 49.07 & 43.84  \\
    
    10   & 59.88 & 51.71  \\
    \hline
    \hline

\end{tabular}
\end{table}
Bi-Sb alloys are narrow band gap semiconductor. Effect of magnetic field on band
structure of this kind of alloy is much more than wide band gap materials. Semimetallic character of Bi arises due to the
small overlap of valence band maximum at T point with conduction band minimum by 0.0184 eV at L point of the 
Brilloiun zone.\cite{31,33} Addition
of Sb to Bi causes $L_s$ and T bands to move down with respect to $L_a$ band. At $x = 0.04$ L band inverts\cite{3,9,15,20}
and at the range of $x\simeq 0.07$ to 0.22 the overlap between the hole T and $L_a$ bands disappears, resulting in semiconducting Bi-Sb alloy. Effect of magnetic field on the energy bands of Bi was theoretically examined by G. A. Baraff.\cite{36} The calculation is based on the two band model of Cohen and Blount, where the properties of conduction band are determined by the presence of a
valence band separated from it by a very small energy gap.\cite{37} Cohen and Blount showed that, this situation leads to a spin mass
equal to the cyclotron mass and the energy levels E(n,s), labelled by orbital quantum number $n = 0, 1, 2..$ and spin quantum number 
$s = \pm$ 1 have a characteristic degeneracy. Baraff extended his calculation by taking into account the effect of other bands on 
the energy levels E(n,s) by using perturbation theory in which two band model plays the role of unperturbed Hamiltonian and 
demonstrated that magnetic field causes lifting of the degeneracy, resulting in increase of $E_g$.\cite{36}
The Bi rich semiconducting $Bi_{1-x}Sb_x$ alloys discussed here possess similar band picture.\cite{28,31} Hence, in accordance with G. A. Baraff it is justified to assume that, observed increase of $E_g$ in presence
of magnetic field for $Bi_{0.90}Sb_{0.10}$ and $Bi_{0.86}Sb_{0.14}$ is due to the lifting of degeneracy of energy levels.\cite{36,37}

	MR of $Bi_{1-x}­Sb_x (x=0.10, 0.14)$ samples is measured at different temperatures in the range of 30-300 K in magnetic
fields upto $\pm 15$ T 
(Figure 2). At 30 K, MR value as high as $325\%$ has been obtained for $Bi_{0.86}Sb_{0.14}$ at 15 T magnetic field without any sign of saturation. However under similar condition direct gap $Bi_{0.90}Sb_{0.10}$ sample shows much higher
value of MR, around $370\%$. Such large value of MR in polycrystalline Bi rich alloys is noteworthy. 
MR effect in Bi and related materials is the so-called ordinary MR, where as mentioned earlier, magnetic field induced Lorentz force plays a major role.\cite{9,30} Ordinary 
MR in a material is generally governed by the parameter $\omega_c\tau$, where cyclotron frequency $\omega_c=\frac{eH}{m^*c}$, $\tau$ (relaxation time) is proportional to mean free path (l) and e is charge of electron; m* is effective carrier mass and c is speed of light.\cite{30} m* is much smaller in Bi
than in conventional metals, $\omega_c$ is about two orders of magnitude larger. In addition, Bi has long l (equivalent to large $\tau$).
The Bi rich Bi-Sb alloys combine all these properties and should depict high MR value. It is needless to say that long l, which 
determines the large value of MR, is directly related to sample quality.\cite{9,8} Temperature dependent 
powder x-ray diffraction data also indicates the high crystalline quality of the polycrystalline samples reported here.\cite{29}
\FloatBarrier
\begin{figure}[h]
\begin{center}
\includegraphics[scale=1.0]{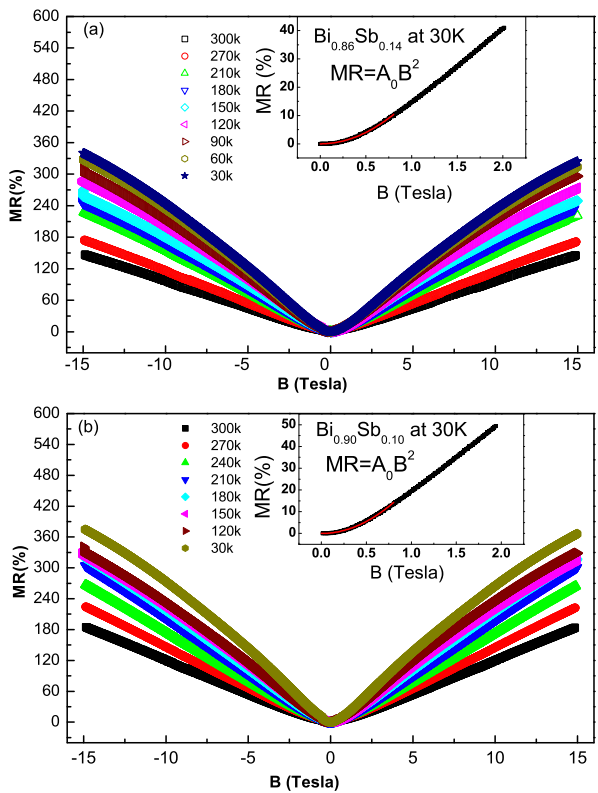}
\caption{(Color online) MR of (a) $Bi_{0.86}Sb_{0.14}$ and (b) $Bi_{0.90}Sb_{0.10}$ alloys at different temperatures. Inset (a) and (b) shows low field MR data at 30K. Solid line indicates fitting with the equation $MR=A_{0}B^{2}$.}
\end{center}
\end{figure}

     It is also observed from Figure 2 that for both samples the value of MR gradually increases with decreasing temperature. 
Acoustic phonon and other temperature dependent scattering decrease with decreasing temperature, causing increase in $\tau$.
Also, electron mobility ($\mu$) should increase at low temperature. Increased $\tau$ and $\mu$ with decreasing temperature 
lead to the observed increase of MR with decreasing temperature. It has been attempted to estimate the temperature dependence of 
$\mu$ from the MR data (given below). Figure 2 further revealed that sample with low Sb content, e.g., $Bi_{0.90}Sb_{0.10}$
possesses higher MR. It is quite expected, since the key in realizing a large MR effect in pristine Bi 
is large value of l, which should decrease with increasing Sb concentration. Figure 2 depicts that, MR first experiences a quadratic growth
below a certain threshold, and then transform into a nearly linear rising with increasing magnetic field without sign of saturation. 
The usual quadratic growth observed at low magnetic field is mainly governed by Kohler's law. According to this simple model, low 
field quadratic MR can be fitted by $MR=A_0B^2$, with $A_0$ a constant.\cite{30} Earlier similar behavior has been observed in single
crystal Bi thin films\cite{9}, single crystals of $Bi_2Se_3$ and $Bi_2Te_3$,\cite{8,15} $Bi_{0.92}Sb_{0.08}$ nanowire.\cite{27} But,
the non saturating MR observed at high magnetic fields should originate from another mechanism, the origin of which is quite
contradictory.\cite{6,8,15} The square root of the prefactor of the quadratic term $(A_0)^\frac{1}{2}$, extracted from fitting low field
MR data [Figure 2 (inset)], is plotted in Figure 3 for both $Bi_{1-x­}Sb_x (x=0.10, 0.14)$ samples.
The unit of $A_0$­ is $T^{-2}(T=Tesla)$, which is equal to the unit of squared mobility $\mu^2$; with 1 $T^{-2}= $1 $m^4V^{-2}s^{-2}$, i.e., the square root of the prefactor term $(A_0)^\frac{1}{2}$ is equal to mobility ($\mu$). Thus temperature dependent behavior of $\mu$ can be extracted from fitting low field MR data 
obtained at different temperatures. Seebeck coefficient measurements confirm that, the carriers are electrons in the
polycrystalline samples reported here.\cite{29} Figure 3 clearly reveals that for both samples $\mu$ increases with decreasing
temperature and $\mu$ of $Bi_{0.90}Sb_{0.10}$ sample is higher. Similar temperature dependence of $\mu$ for single-crystal
Bi and Bi-Sb nanowire was also reported and the present estimated dependences are in good agreement with this trend.\cite{27,33}
As $\mu$ is suppressed by scattering at grain boundaries, its magnitude in polycrystalline Bi-Sb alloy will
be always lower than that of Bi single crystal.\cite{12} Further, with increasing Sb concentration number of scattering sites increases,
this tends to decrease $\mu$ in higher Sb content sample e.g., $Bi_{0.86}Sb_{0.14}$. Temperature dependence of $\mu$ for polycrystalline $Bi_{1-x}Sb_x (x = 0.10, 0.14)$ samples has been estimated. For pristine Bi,
$\mu$ is proportional to $T^m$ [11]. The value of `m' 
for polycrystalline Bi-Sb alloys is extracted by fitting $\mu$-T data. The obtained `m' values are 1.45 and 1.53, respectively 
for $Bi_{0.90}Sb_{0.10}$ and $Bi_{0.86}Sb_{0.14}$ alloys. For single crystal Bi samples, `m' values of similar magnitude was reported by Hasegawa et al.\cite{12} Temperature dependence of MR value obtained at 15 T is also co-plotted in Figure 3 and observed that, there is some correlation between MR and $\mu$ values and they follow similar temperature dependence. The small hump observed around 180 K in the MR-T curve might be related to the metal-semiconductor transition depicted in Figure 1. This observation might unveil interesting feature to be studied carefully.
\FloatBarrier
\begin{figure}[h]
\begin{center}
\includegraphics[scale=0.9]{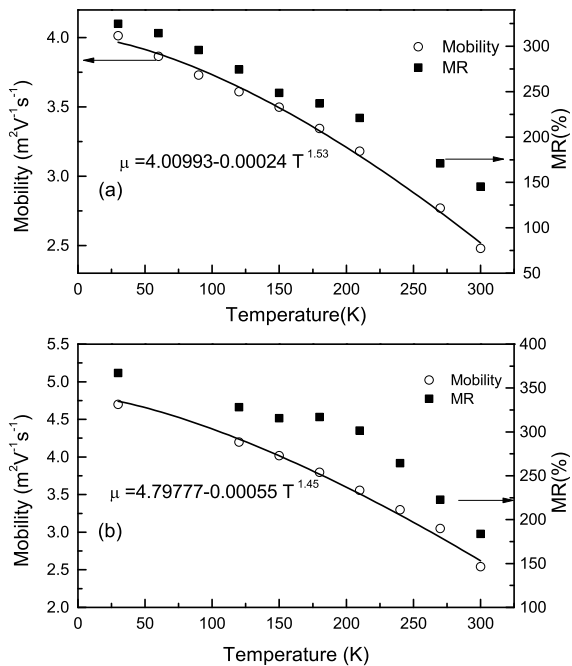}
\caption{Temperature dependent mobility and MR of (a) $Bi_{0.86}Sb_{0.14}$ and (b) $Bi_{0.90}Sb_{0.10}$ alloys. Solid line indicates the fitting of temperature dependent mobility ($\mu$) of the samples with the equation: $\mu=a+bT^m$.}
\end{center}
\end{figure}

  The well established Kohler's law suggests that, MR is a universal function of B and unlike low field quadratic behavior observed 
upto around 1 T, at high fields MR will saturate in most materials. Thus, such a giant, nearly linear, non saturating MR observed in these polycrystalline samples are interesting. The on-set field $(B^*)$ of linear behavior observed here is around 1 Tesla 
(Figure 2 (inset)). Very recently, Wang et al. also reported similar crossover at a critical field $(B^*)$ from semiclassical $B^2$ dependence to the high field LMR for $Sb_2Te_2Se$ system.\cite{42} A general model for LMR at strong field is proposed by Abrikosov, in which all the electrons are populated into
lowest Landau level when ``extreme quantum limit'' $\hbar\omega_c>E_F$ is satisfied, where E$_{F}$ is Fermi energy. \cite{39} A more accurate inequality $n_0<<(\frac{eH}{\hbar c})^{\frac{3}{2}}$ is provided to estimate $B^*$, where symbols have their usual meaning.\cite{8} In our case carrier concentration, estimated using $\mu$ value, is around $6\times10^{17}$ $cm^{-3}$. This leads B* to be larger than 10 T. B* with same order of 
magnitude are also reported for $Bi_2Se_3$ nanoplates, thin films and single crystalline $Sb_2Te_2Se$ system.\cite{8,17,42} Whereas Zhang et al reported much higher B* value in $Bi_2Te_3$ thin films.\cite{6} Experimentally observed B* in our samples 
is much lower ($\sim 1$T), thus we conclude the Abrikosov quantum model may not be applicable here.

	By using classical method, Paris and Littlewood proposed a model (P-L model) that solves the mystery of LMR found
in disordered systems.\cite{40,41} This classical model predicts that B* can be estimated by the relation $B^*=\mu^{-1}$. From the estimated value of $\mu$ (Figure 3) B* is of the order of 1 T and corroborate 
with the MR curves [Figure 2(inset)]. Further the consistent behavior of MR and $\mu$ throughout the measured temperature 
range from 30 K to 300 K [Figure 3] gives direct evidence in support of applicability of P-L model in $Bi_{1-x}Sb_{x} (x=0.10, 0.14)$
alloys.\cite{8} Additionally, according to this model $\mu$ dominates the evaluation of MR if $\frac{\Delta\mu}{\mu}<1$
 , i.e., spatial distribution of mobility is overwhelmed by mobility itself. This is satisfied for these high quality polycrystalline Bi-Sb samples with a remarkable mobility of 4.7 $m^2V^{-1}s^{-1}$. Recently, Yan et al also reported a clear evidence of applicability of P-L model in their $Bi_2Se_3$ samples.\cite{8}

  In summary, we observed non saturating (nearly linear) MR in thermoelectric $Bi_{1-x}Sb_x (x=0.10, 0.14)$ alloys. Unusual,
giant MR of around $400\%$ at 30 K and $150\%$ at 300 K has been reported. Such large MR indicates the high quality of the 
sample and will be valuable for magnetic field sensors over a broad temperature range. Under the influence of magnetic field, E$_g$ increases which might be related to the lifting of the degeneracy under 
the application of magnetic field. Low field MR data follows Kohler's law (quadratic growth), where conduction electrons are deflected by the Lorentz force in presence of magnetic field. However, with increasing magnetic field MR depicts more or less linear behavior. Temperature dependence of $\mu$, extracted from low field MR data, follow similar trend as that of carriers in Bi. Both MR and $\mu$ are high for $Bi_{0.90}Sb_{0.10}$ sample. Proportional relation between $\mu$ and MR has been observed. It has been demonstrated that, nearly linear non saturating high field MR is of classical origin and evidence has been given in support of the classical P-L model.

  We acknowledge DST, Govt. of India for sanctioning research project[$SR/FTP/PS-25/2009$] and UGC DAE CSR, Kolkata, for providing the 15 T magnetoresistance measurement facility. DD and KM is greatful to DST and UGC for providing their Research Fellowship.

\end{document}